\documentclass[12pt]{article}
\usepackage{uereport}
\usepackage{verbatim}
\usepackage{siunitx}
\usepackage{url}
\usepackage[breaklinks]{hyperref}
\usepackage{graphicx}
\usepackage[margin=1cm,small]{caption}
\usepackage{subcaption}
\usepackage{mathrsfs} 
\usepackage{amsmath}
\usepackage{nicefrac}
\usepackage{lipsum}
\usepackage[defaultlines=3,all]{nowidow}

\title{Wireless Power Transfer by Means of Electromagnetic Radiation Within an\\Enclosed Space}
\author{Robert A. Moffatt\\Stanford University Department of Physics\\\href{mailto:rmoffatt@stanford.edu}{rmoffatt@stanford.edu}}
\course{\today}

\begin{document}
\maketitle

\abstract{In this paper, wireless power transfer by means of electromagnetic radiation is investigated. Formulas are derived for the efficiency of the power transfer in free space, in the presence of reflecting surfaces, and within enclosed spaces. It is found that the presence of reflecting surfaces has the capacity to substantially enhance the efficiency of power transfer at long range. An upper limit is also found for the transferred power when constraints are imposed on certain forms of undesired absorption. For the sake of simplicity, only the efficiency of the radiative power transfer is considered. Losses due to resistance in the antenna structures or inefficiencies in RF to DC conversion are neglected.}

\section{Introduction}

The concept of wireless power transfer has a long history, beginning with the research of Nikola Tesla over a century ago \cite{Tesla1914,Brown1984}. Since the late 1950's, beamed microwaves  have been investigated for their possible use in transferring power to or from satellites in orbit \cite{Brown1984,Brown1992}. Perhaps the most impressive demonstration of long-distance wireless power transfer by beamed microwaves was a test performed by NASA at the JPL Goldstone Facility in 1975, in which 30kW of power were transferred over a distance of one mile with a DC-to-DC efficiency of over 50\% \cite{Dickinson1975a,Dickinson1975b}.

Recent decades have seen a proliferation of small, battery-powered consumer electronic devices. As a result, there has been a renewed interest in wireless power specifically for the purpose of charging such devices within a domestic setting \cite{Kurs2007,Gowda2016,Smith2016}. Compared to free space, a domestic setting differs in three important ways: First, unlike a receiver in free space, a receiver in a domestic setting will typically be within the Fresnel region of the transmitter, allowing certain focusing and beam-forming techniques to be employed \cite{Gowda2016,Smith2016}. Second, human beings will typically be present near the transmitter and/or receiver, so safety limits on human exposure will place strict limitations on the maximum power density which can exist in the vicinity of either of these devices. And third, many reflecting surfaces will typically be present in a domestic setting which can significantly alter the propagation of radiation from the transmitter to the receiver.

The last difference is the subject of recent research which suggests that these reflections may substantially enhance the efficiency of radiative power transfer in a domestic setting over that which would be possible across similar distances in free space \cite{Leabman2014}. In order to further explore this possibility, both the efficiency and the safety limitations of radiative power transfer within an enclosed space are investigated in the following sections.

\section{A Diffraction-Limited Beam in Free Space}

For the sake of comparison, let us first review the efficiency of beamed power in free space \cite{Brown1984,Brown1992,Gowda2016,Smith2016}. Consider a receiver which receives electromagnetic power from a transmitter. Let it be assumed that the transmitter has a circular aperture of diameter, $d$, and produces a diffraction-limited beam focused on the position of the receiver, which is a distance, $z$, away from the transmitter along its optical axis.

\begin{figure}[htb]
\begin{center}
\includegraphics[width=\textwidth]{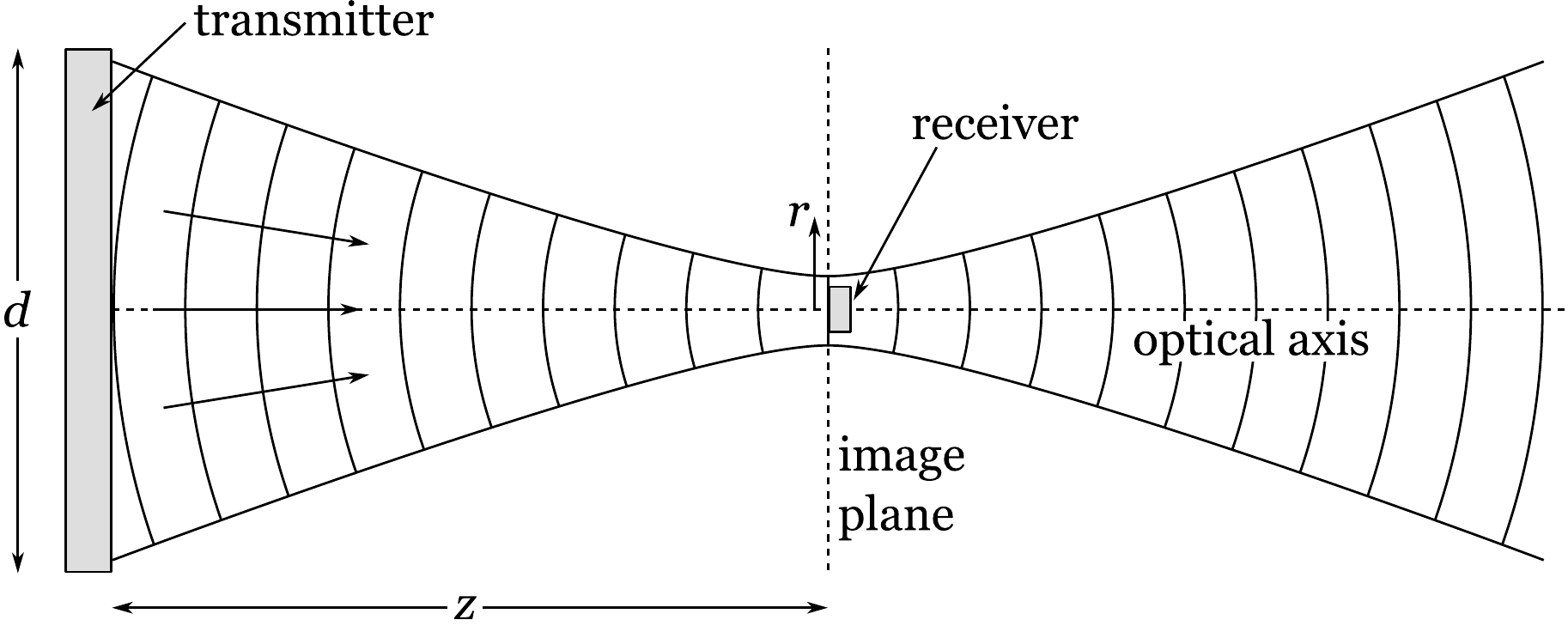}
\caption{A diffraction-limited beam generated by a circular transmitter antenna of diameter, $d$. The beam is focused on a receiver at a distance $z$ from the transmitter.}
\label{fig:diffraction_limited_beam}
\end{center}
\end{figure}

Let the coordinate, $r$, denote the distance from the optical axis. Let the image plane be defined to be the plane perpendicular to the optical axis which passes through the location of the receiver. The intensity of radiation in the image plane, $I(r)$, is given by \cite{Jackson}:
\begin{equation}
I(r) = I_0 \left( \frac{2 J_1\left( \pi \frac{d \, r}{z \lambda} \right)}{\pi \frac{d \, r}{z \lambda}} \right)^2
\end{equation}
where $J_1$ is the first-order Bessel function of the first kind, $\lambda$ is the wavelength of the radiation, and $I_0$ is the intensity at the center, given by:
\begin{equation}
I_0 = P_T \frac{\pi d^2}{4 z^2 \lambda^2}
\label{eqn:intensity}
\end{equation}
where $P_T$ is the transmitted power.

Assuming both antennas have the same polarization, the power received by the receiver, $P_R$, is given by:
\begin{equation}
P_R = \mathcal{A}_R I_0
\end{equation}
where $\mathcal{A}_R$ is the effective area of the receiving antenna. It is assumed that the receiver antenna is exactly centered in the beam of radiation emanating from the transmitter, and that the receiver is small compared to the width of the beam. The effective area, $\mathcal{A}$, of an antenna is related to its gain, $G$, by the equation \cite{Balanis}:
\begin{equation}
\mathcal{A} = \frac{\lambda^2}{4 \pi} G
\label{eqn:area_and_gain}
\end{equation}
Therefore, the efficiency, $\eta$, of the power transfer is\footnote{Note that equation \ref{eqn:efficiency} is only accurate when $\mathcal{A}_R \mathcal{A}_T$ is much smaller than $\lambda^2 z^2$. Reference \cite{Brown1992} provides a plot of the efficiency, $\eta$, as a function of $\tau = \sqrt{\mathcal{A}_R \mathcal{A}_T}/(\lambda z)$ which is accurate for all values of $\tau$.}:
\begin{equation}
\eta = \frac{P_R}{P_T} = \mathcal{A}_R \frac{\pi d^2}{4 z^2 \lambda^2} = \frac{G_R}{4\pi} \frac{\pi (d/2)^2}{z^2} = \frac{G_R \mathcal{A}_T}{4 \pi z^2} = \frac{\mathcal{A}_R \mathcal{A}_T}{\lambda^2 z^2}
\label{eqn:efficiency}
\end{equation}
where $G_R$ is the gain of the receiver antenna, and $\mathcal{A}_T$ is the effective area of the transmitter antenna, assuming that $d$ is much greater than the wavelength.

\section{Reciprocity}

Equation \ref{eqn:efficiency} may also be understood from a different perspective, which provides some insight into the operation of this system. According to the Electromagnetic Reciprocity Theorem \cite{Balanis}, the efficiency of power transfer between two antennas is unchanged if the roles of the transmitter and receiver are exchanged.

\begin{figure}[htb]
\begin{center}
\includegraphics[width=0.7\textwidth]{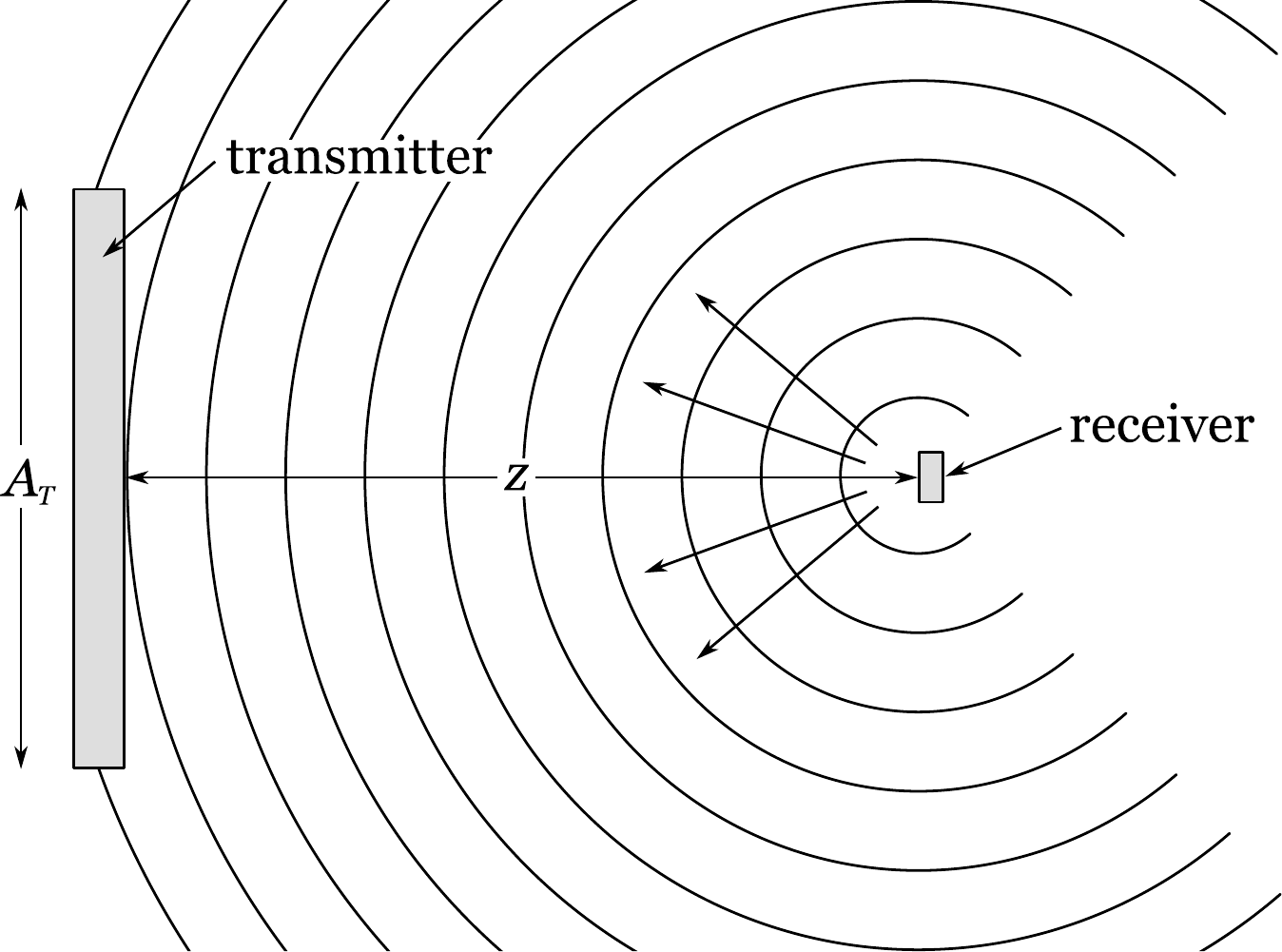}
\caption{Power sent in reverse from the receiver to the transmitter. Here, the receiver antenna is shown with gain greater than 1 in the direction of the transmitter antenna, and gain less than 1 in the opposite direction. The receiver is a distance $z$ from the transmitter, and the cross-sectional area of the transmitter antenna is $\mathcal{A}_T$.}
\label{fig:reciprocity}
\end{center}
\end{figure}

Consider the case in which a power, $P_R$, is sent in reverse from the receiver antenna back to the transmitter antenna. In the vicinity of the transmitter antenna, the power density, $S$, is:
\begin{equation}
S = P_R \frac{G_R}{4 \pi z^2}
\label{eqn:power_density}
\end{equation}
Assuming the transmitter antenna collects all of the radiation incident on its aperture, the power, $P_T$, received by the transmitter antenna will be:
\begin{equation}
P_T = \mathcal{A}_T S
\end{equation}
Here, it is assumed that the power density, $S$, is approximately uniform across the aperture of the transmitter antenna. The efficiency, $\eta$, of the power transfer is:
\begin{equation}
\eta = \frac{P_T}{P_R} = \frac{G_R \mathcal{A}_T}{4 \pi z^2}
\label{eqn:efficiency_general}
\end{equation}
which is the same as equation \ref{eqn:efficiency}, except that it may be applied in the general case of a transmitter antenna of arbitrary shape. Note that if the receiver is off-axis, the area of the transmitter antenna, $\mathcal{A}_T$, must be replaced by its effective area in the direction of the receiver.

\section{The Effect of Reflections}
\label{sec:reflections}

Let us now explore how the efficiency of power transfer is affected in the vicinity of a reflecting object. Consider a transmitter-receiver pair located near an infinite plane exhibiting specular reflection with power reflection coefficient, $\rho$. Unlike the arrangement shown in Figure \ref{fig:diffraction_limited_beam}, there are two possible diffraction-limited beams which can deliver power from the transmitter to the receiver: a direct beam, and a reflected beam. 

\begin{figure}[htb]
\begin{center}
\includegraphics[width=\textwidth]{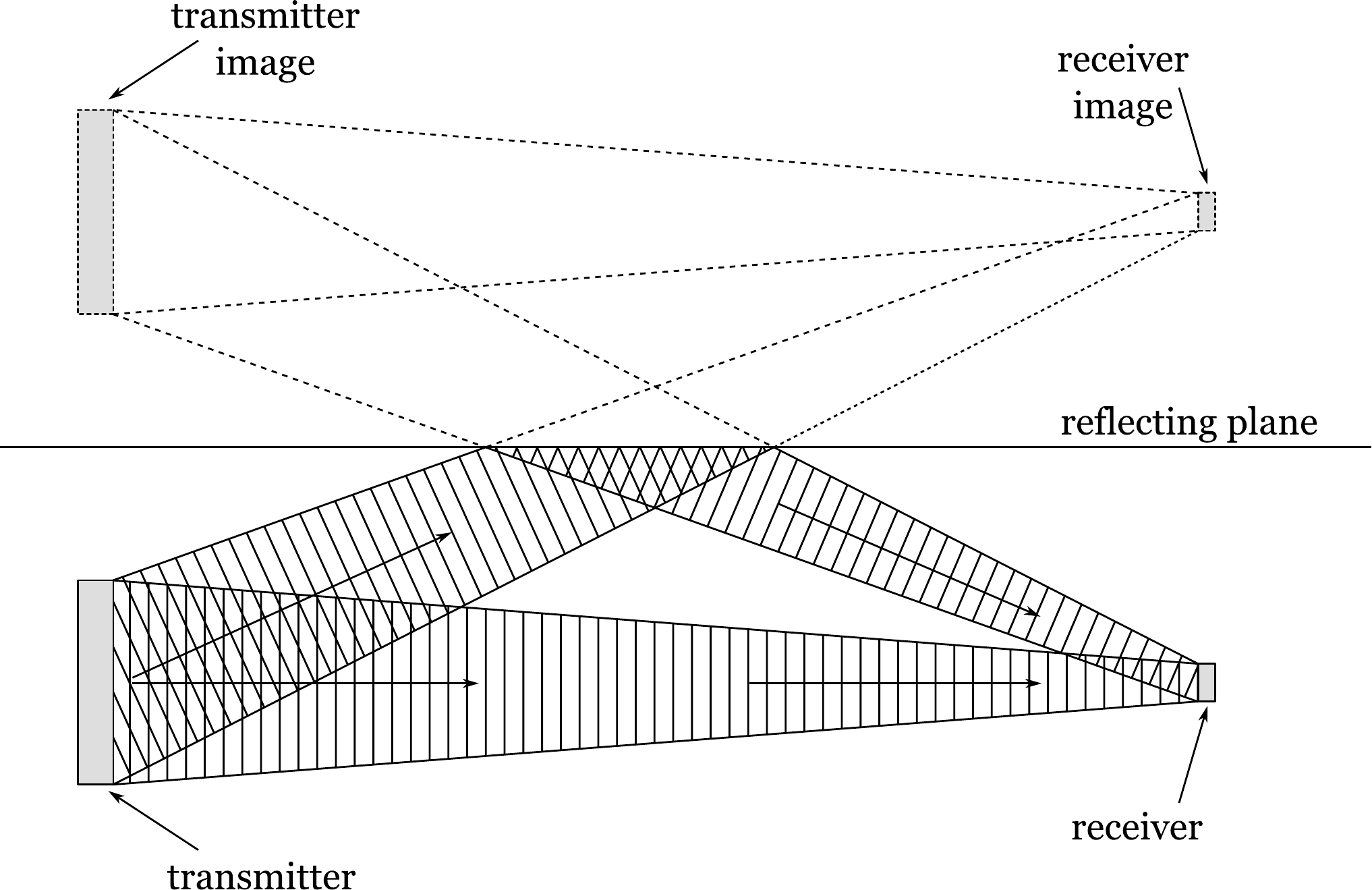}
\caption{A transmitter-receiver pair in the vicinity of a reflecting plane.}
\label{fig:single_reflection}
\end{center}
\end{figure}

Let $\eta_1$ be the efficiency of the direct beam, and let $\eta_2$ be the efficiency of the reflected beam. Using equation \ref{eqn:efficiency_general}, we can write these efficiencies as:
\begin{eqnarray}
\eta_1 &=& \frac{G_{R1} \mathcal{A}_{T1}}{4 \pi z_1^2}\nonumber\\
\eta_2 &=& \frac{G_{R2} \mathcal{A}_{T2}}{4 \pi z_2^2} \rho
\end{eqnarray}
where $G_{R1/2}$ is the gain of the receiver antenna in the direction of the incoming direct/reflected beam, $z_{1/2}$ is the path length of the direct/reflected beam, and $\mathcal{A}_{T1/2}$ is the effective area of the transmitter antenna in the direction of the outgoing direct/reflected beam.

Now consider the case in which the transmitter antenna emits both beams simultaneously\footnote{Such an arrangement is possible, for example, if the transmitter antenna consists of a phased array of patch antennas, where the amplitude and phase of each antenna are able to be controlled independently \cite{Smith2016,Leabman2014}.}, with a fraction of power $f_1$ allocated to the direct beam, and a fraction $f_2=1-f_1$ allocated to the reflected beam. The amplitude of the direct beam is proportional to $\sqrt{f_1 \eta_1}$, and the amplitude of the reflected beam is proportional to $\sqrt{f_2 \eta_2}$. The overall efficiency, $\eta$, is equal to the squared magnitude of the complex sum of these two amplitudes:
\begin{equation}
\eta = f_1 \eta_1 + f_2 \eta_2 + 2 \sqrt{f_1 \eta_1 f_2 \eta_2} \cos{\phi}
\label{eqn:reflection_efficiency}
\end{equation}
where $\phi$ is the phase difference between the two beams at the location of the receiver.\footnote{Here we assume that the receiver is small enough that the spatial variation in the phase of each beam may be neglected.} This efficiency is maximized when both beams are in phase at the location of the receiver:
\begin{equation}
\phi = 0
\end{equation}
and when the fraction of power allocated to each beam is given by:
\begin{eqnarray}
f_1 &=& \frac{\eta_1}{\eta_1+\eta_2}\nonumber\\
f_2 &=& \frac{\eta_2}{\eta_1+\eta_2}
\label{eqn:power_fraction}
\end{eqnarray}
The maximized efficiency is:
\begin{equation}
\eta_{\textrm{max}} = \eta_1+\eta_2
\end{equation}

In contrast, consider the case in which the two beams are mutually incoherent. The incoherent efficiency, $\eta_{\textrm{inc}}$, is given by the average of equation \ref{eqn:reflection_efficiency} over all phase angles, $\phi$:
\begin{equation}
\eta_{\textrm{inc}} = f_1 \eta_1 + f_2 \eta_2
\label{eqn:incoherent_efficiency}
\end{equation}
From equation \ref{eqn:incoherent_efficiency}, we see that the incoherent efficiency is strictly less than or equal to the greater of the two efficiencies, $\eta_1$ or $\eta_2$.

Therefore, we see that the presence of a reflecting surface allows the efficiency of power transfer to be enhanced, but only if the phases of the wavefronts leaving the transmitter are controlled so as to produce constructive interference in the vicinity of the receiver.

\section{Reflections and Reciprocity}
\label{sec:reflections_and_reciprocity}

\begin{figure}[htb]
\begin{center}
\includegraphics[width=\textwidth]{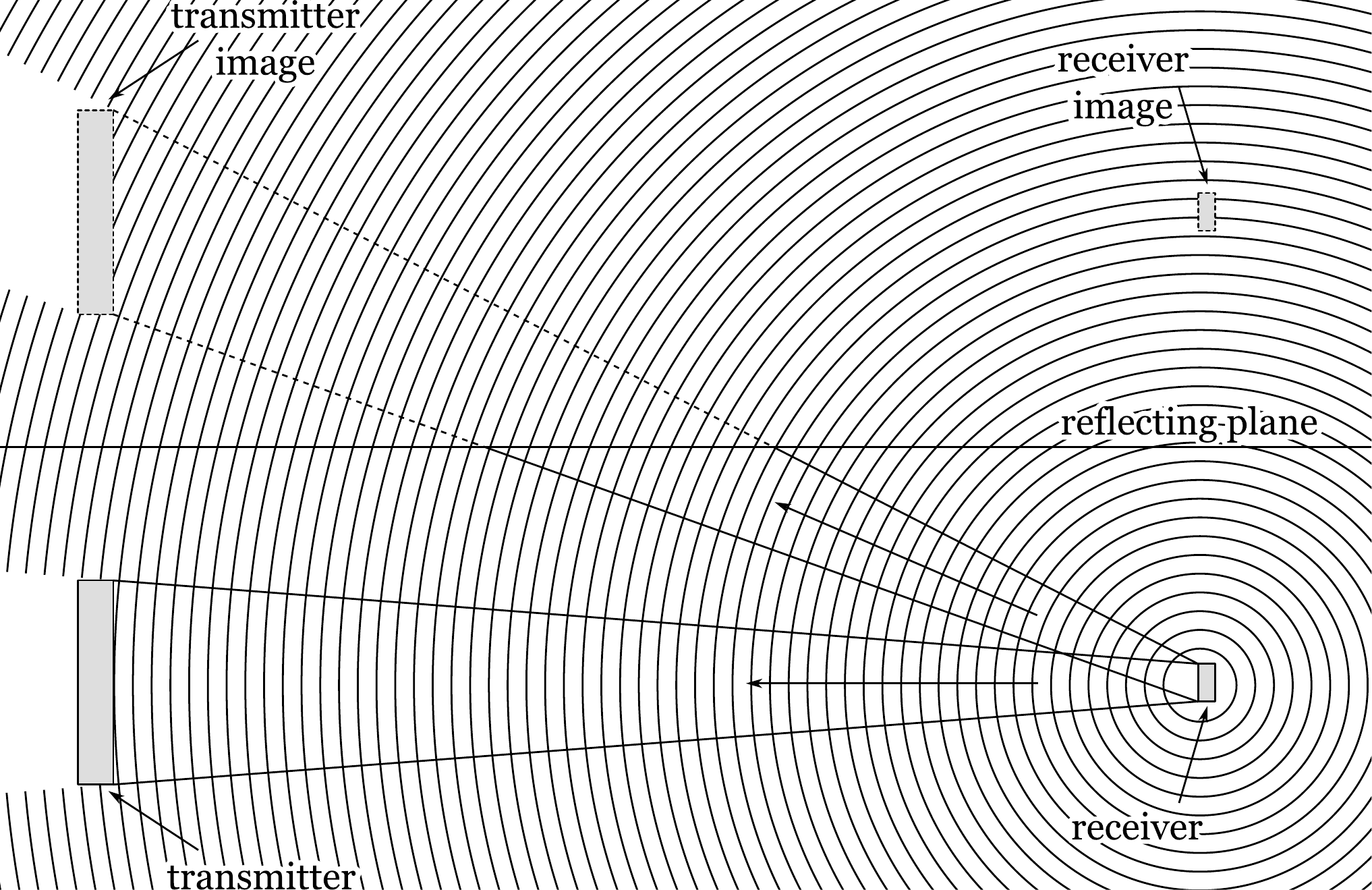}
\caption{Power sent in reverse from the receiver to the transmitter in the vicinity of a reflecting plane. Power is absorbed by both the transmitter and its image.}
\label{fig:reflection_reciprocity}
\end{center}
\end{figure}

The result of section \ref{sec:reflections} may also be derived using reciprocity, in which the roles of the receiver and transmitter are reversed. Consider the arrangement shown in Figure \ref{fig:reflection_reciprocity}, which is the same as that shown in Figure \ref{fig:single_reflection}, except that power is sent in the opposite direction from the receiver back to the transmitter.

The power radiates outward from the receiver and is absorbed by both the transmitter and its image. Assuming that the receiver emits power $P_R$, the power density, $S_1$, incident on the transmitter is:
\begin{equation}
S_1 = P_R \frac{G_{R1}}{4 \pi z_1^2}
\end{equation}
and the power density, $S_2$, incident on the image of the transmitter is:
\begin{equation}
S_2 = P_R \frac{G_{R2}}{4 \pi z_2^2} \rho
\end{equation}
where $G_{R1/2}$ and $z_{1/2}$ are defined in the same manner as before. The power absorbed by the transmitter is:
\begin{equation}
P_{T1} = \mathcal{A}_{T1} S_1
\end{equation}
and the power absorbed by the image of the transmitter is:
\begin{equation}
P_{T2} = \mathcal{A}_{T2} S_2
\end{equation}
where, as before, $\mathcal{A}_{T1/2}$ is the effective area which the transmitter presents to the incident radiation along the path of the direct/reflected beam. The overall efficiency, $\eta$, is:
\begin{equation}
\eta = \frac{P_{T1}+P_{T2}}{P_R} =  \frac{G_{R1} \mathcal{A}_{T1}}{4 \pi z_1^2} +  \frac{G_{R2} \mathcal{A}_{T2}}{4 \pi z_2^2} \rho = \eta_1 + \eta_2
\end{equation}
This is the same as the maximum efficiency derived in section \ref{sec:reflections}, except that the methodology used in this section can be more easily applied to general arrangements of the transmitter and receiver.

\section{A Cubical Room with Reflective Walls}
\label{sec:cubical_room}

Having shown that the presence of a reflecting plane allows the efficiency of wireless power transfer to be enhanced, we next wish to explore a situation in which both the transmitter and receiver are completely enclosed by reflecting walls.

Consider a transmitter and receiver inside a cubical room with side length, $h$. Assume that the transmitter is a sphere with surface area $\mathcal{A}_T$, covered by an array of small antennas, and centered within the cubical room. Assume that the walls of the room are smooth, and exhibit specular reflection, with power reflection coefficient, $\rho$, and power absorption coefficient, $\alpha = 1-\rho$.

\begin{figure}[htb]
\begin{center}
\includegraphics[width=0.6\textwidth]{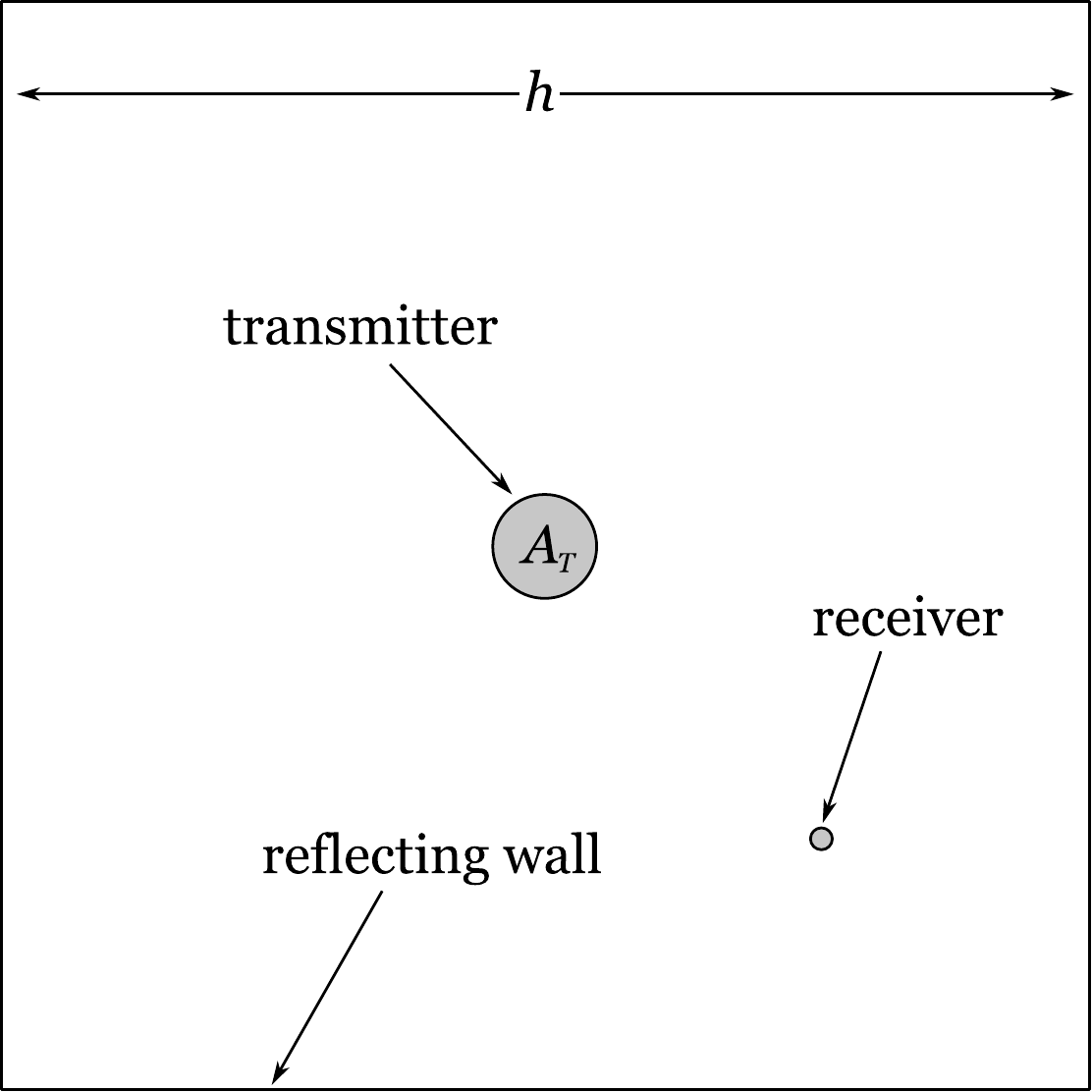}
\caption{A transmitter and receiver in a cubical room with side length $h$, viewed from above. The transmitter is a sphere with surface area $\mathcal{A}_T$, located at the center of the room.}
\label{fig:cubical_room}
\end{center}
\end{figure}

Using reciprocity, we may treat the receiver as the source of RF power. We wish to find the fraction of this power which is absorbed by the transmitter. This problem is complicated by the fact that the radiation can undergo any number of reflections before it reaches the surface of the transmitter.

\begin{figure}[htb]
\begin{center}
\includegraphics[width=0.8\textwidth]{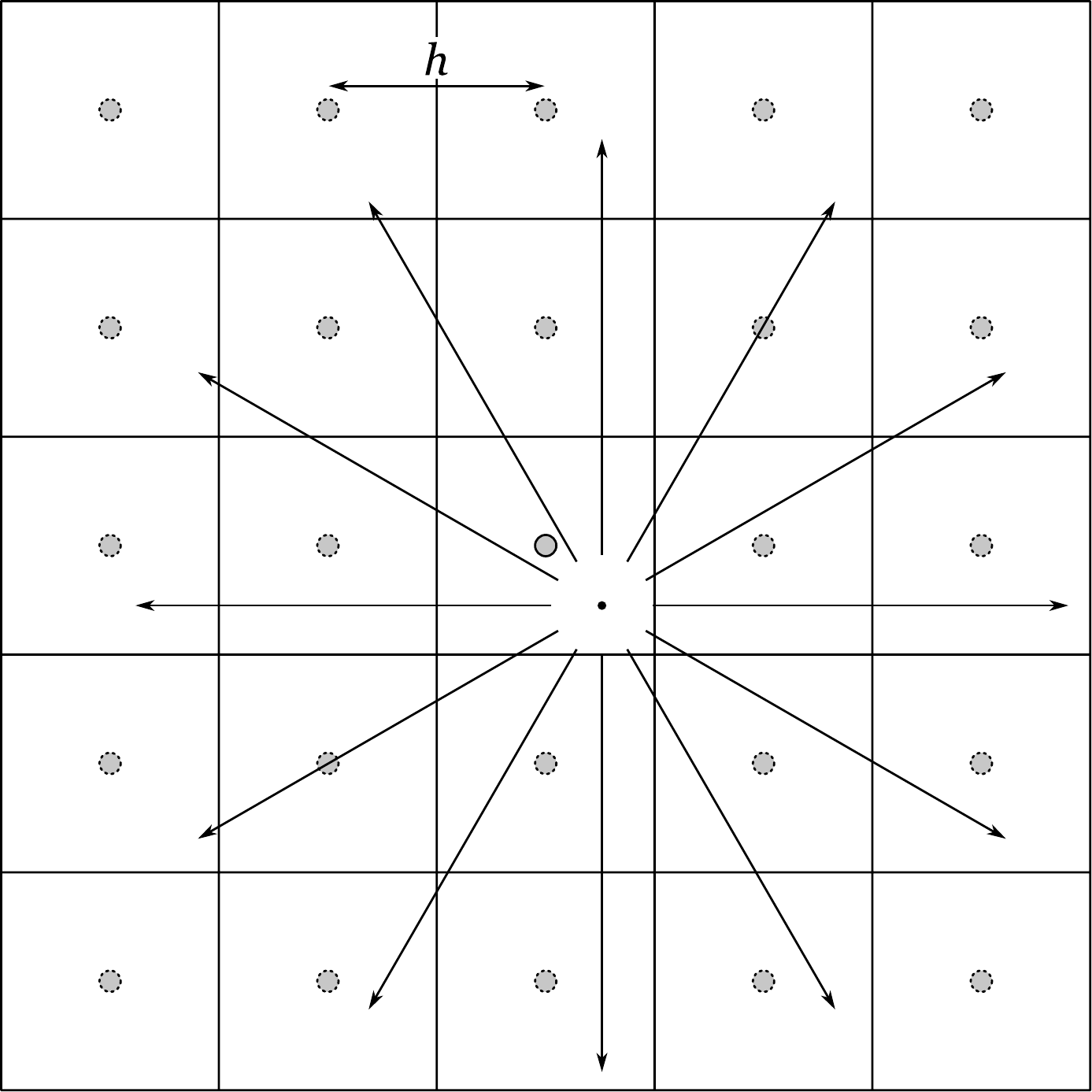}
\caption{Images of the transmitter formed by the reflecting walls of the room. The radiation spreads out radially away from the receiver in the image space.}
\label{fig:cubical_room_images}
\end{center}
\end{figure}

The problem can be better understood in terms of images, as shown in Figure \ref{fig:cubical_room_images}. The images of the transmitter form an infinite cubical grid, with lattice constant $h$. In the image space, the walls may be treated as partially transparent, with an absorption coefficient equal to $\alpha$ as before, a transmission coefficient of $\tau = 1-\alpha$, and a reflection coefficient of zero. The efficiency of power transfer is therefore equal to the fraction of the radiation emanating from the receiver which is absorbed by the transmitter and its infinite lattice of images.

The fraction of power absorbed by the transmitter directly, $\eta_{\textrm{direct}}$, is given by:
\begin{equation}
\eta_{\textrm{direct}} = \frac{G_R \, \sigma}{4\pi r^2}
\label{eqn:direct_efficiency}
\end{equation}
where $r$ is the distance from the transmitter to the receiver, $\sigma = \mathcal{A}_T/4$ is the cross-sectional area of the transmitter antenna, and $G_R$ is the gain of the receiver in the direction of the transmitter.

The rest of the power will spread throughout the image space. Let us assume that both the absorption coefficient, $\alpha$, and the surface area, $\mathcal{A}_T$, are small, such that this radiation can undergo many reflections before being absorbed. In this case, the radiation in the image space will travel many times the lattice constant, $h$. Because the images of the transmitter are distributed omnidirectionally around the receiver, there is no benefit to using a receiver antenna with a gain substantially greater than one\footnote{assuming that the efficiency of the direct path is minimal}. To further simplify the calculation, let us therefore find the efficiency of power transfer averaged over all possible receiver orientations. The receiver gain, $G_R$, may be replaced by its rotational average, which is equal to one for an ideal, lossless antenna. The average of $\eta_{\textrm{direct}}$ over all possible orientations of the receiver is:
\begin{equation}
\overline{\eta}_{\textrm{direct}} = \frac{\sigma}{4\pi r^2}
\end{equation}

The fraction of power not absorbed directly is $1-\eta_{\textrm{direct}}$. Assuming that the transmitter does not subtend a significant fraction of the solid angle as seen from the location of the receiver, the radiation which is not directly absorbed will spread through the image space in all directions. As the radiation reaches distances much larger than the lattice constant, $h$, the wavefronts passing through each unit cell approach more and more closely the form of plane waves. As the radiation passes through each unit cell, some fraction is absorbed by the walls, and some fraction is absorbed by the image of the transmitter. Because these are the only structures which absorb radiation\footnote{Here we neglect the absorption of radiation by images of the receiver. The receiver is assumed to have negligible cross-section relative to the transmitter.}, the fraction of the reflected power absorbed by all of the images of the transmitter is equal to the fraction of radiation absorbed by the transmitter in a unit cell exposed to a plane wave, averaged over all possible directions of the incident plane wave.

If the plane wave has power density, $S$, the power absorbed by the transmitter, $P_T$, is:
\begin{equation}
P_T = \sigma S
\end{equation}
Because the cross-section, $\sigma$, of the spherical antenna is independent of direction, $P_T$ is also equal to the power received by the transmitter averaged over all directions, $\overline{P}_T$:
\begin{equation}
\overline{P}_T = P_T = \sigma S = \mathcal{A}_T S /4
\end{equation}
The power absorbed by one of the walls of the room is:
\begin{equation}
P_{\textrm{one wall}} = h^2 \alpha |\cos{\theta}| S
\end{equation}
where $\theta$ is the angle of incidence of the plane wave relative to the wall. Averaged over all directions, this power is:
\begin{equation}
\overline{P}_{\textrm{one wall}} = h^2 \alpha S/2
\end{equation}
Each unit cell contains three walls. Therefore, the average power, $\overline{P}_W$, absorbed by the walls of the unit cell is:
\begin{equation}
\overline{P}_W = 3 \overline{P}_{\textrm{one wall}} = 3 h^2 \alpha S/2 = \mathcal{A}_W \alpha S/4
\end{equation}
where $\mathcal{A}_W = 6 h^2$ is the total surface area of all six walls in the room. The fraction of the reflected power absorbed by the transmitter, averaged over all orientations of the receiver, $\overline{\eta}_{\textrm{refl.}}$, is therefore:
\begin{equation}
\overline{\eta}_{\textrm{refl.}} = \frac{\overline{P}_T}{\overline{P}_W + \overline{P}_T} = \frac{\mathcal{A}_T}{\mathcal{A}_W \alpha + \mathcal{A}_T}
\label{eqn:box_reflection_efficiency}
\end{equation}
and the overall efficiency, $\overline{\eta}_{\textrm{room}}$, averaged over all orientations of the receiver, is:
\begin{equation}
\overline{\eta}_{\textrm{room}} = \overline{\eta}_{\textrm{direct}} + (1- \overline{\eta}_{\textrm{direct}}) \overline{\eta}_{\textrm{refl.}} = \overline{\eta}_{\textrm{direct}} + \overline{\eta}_{\textrm{refl.}} - \overline{\eta}_{\textrm{direct}} \overline{\eta}_{\textrm{refl.}}
\label{eqn:combined_efficiency}
\end{equation}

We see that at sufficiently large separations between the transmitter and receiver, $\overline{\eta}_{\textrm{direct}}$ becomes negligible, and the efficiency, $\overline{\eta}_{\textrm{room}}$, becomes independent of the position of the receiver. In this limit, the efficiency, $\overline{\eta}_{\textrm{room}}$, is equal to the ratio of the surface area of the transmitter antenna, $\mathcal{A}_T$, divided by the sum of its surface area and the surface area of the walls, $\mathcal{A}_W$, weighted by the absorption coefficient, $\alpha$.

Note that as $\alpha$ approaches 0, the efficiency, $\overline{\eta}_{\textrm{room}}$, approaches 100\%. This is expected, as the transmitter antenna is the only structure capable of absorbing power in this limit. In the limit where $\mathcal{A}_W \alpha \gg \mathcal{A}_T$ and $r$ is of the same order as $h$, the efficiency, $\overline{\eta}_{\textrm{room}}$, becomes:
\begin{equation}
\overline{\eta}_{\textrm{room}} \approx \overline{\eta}_{\textrm{refl.}} \approx \frac{\mathcal{A}_T}{\mathcal{A}_W \alpha} = \frac{2 \sigma}{3 h^2 \alpha} \ \ , \ \ \mathcal{A}_W \alpha \gg \mathcal{A}_T
\end{equation}
The efficiency of a diffraction-limited beam in free space for a receiver at a distance of $\sqrt{3} h/2$, averaged over all receiver orientations, is:
\begin{equation}
\overline{\eta}_{\textrm{free space}} = \frac{\sigma}{4 \pi (\sqrt{3} h/2)^2} = \frac{\sigma}{3 \pi h^2}
\end{equation}
For a receiver located in one of the corners of the room, the presence of the reflecting walls enhances the efficiency by the factor:
\begin{equation}
\frac{\overline{\eta}_{\textrm{room}}}{\overline{\eta}_{\textrm{free space}}} \approx \frac{2 \pi}{\alpha}
\end{equation}
We therefore see that if $\alpha$ is small, the presence of the reflecting walls can provide a great enhancement in the efficiency of the power transfer relative to the maximum efficiency achievable at similar distances in free space.

\section{A Room of Arbitrary Shape}
\label{sec:arbitrary_room}

\begin{figure}[htb]
\begin{center}
\includegraphics[width=0.8\textwidth]{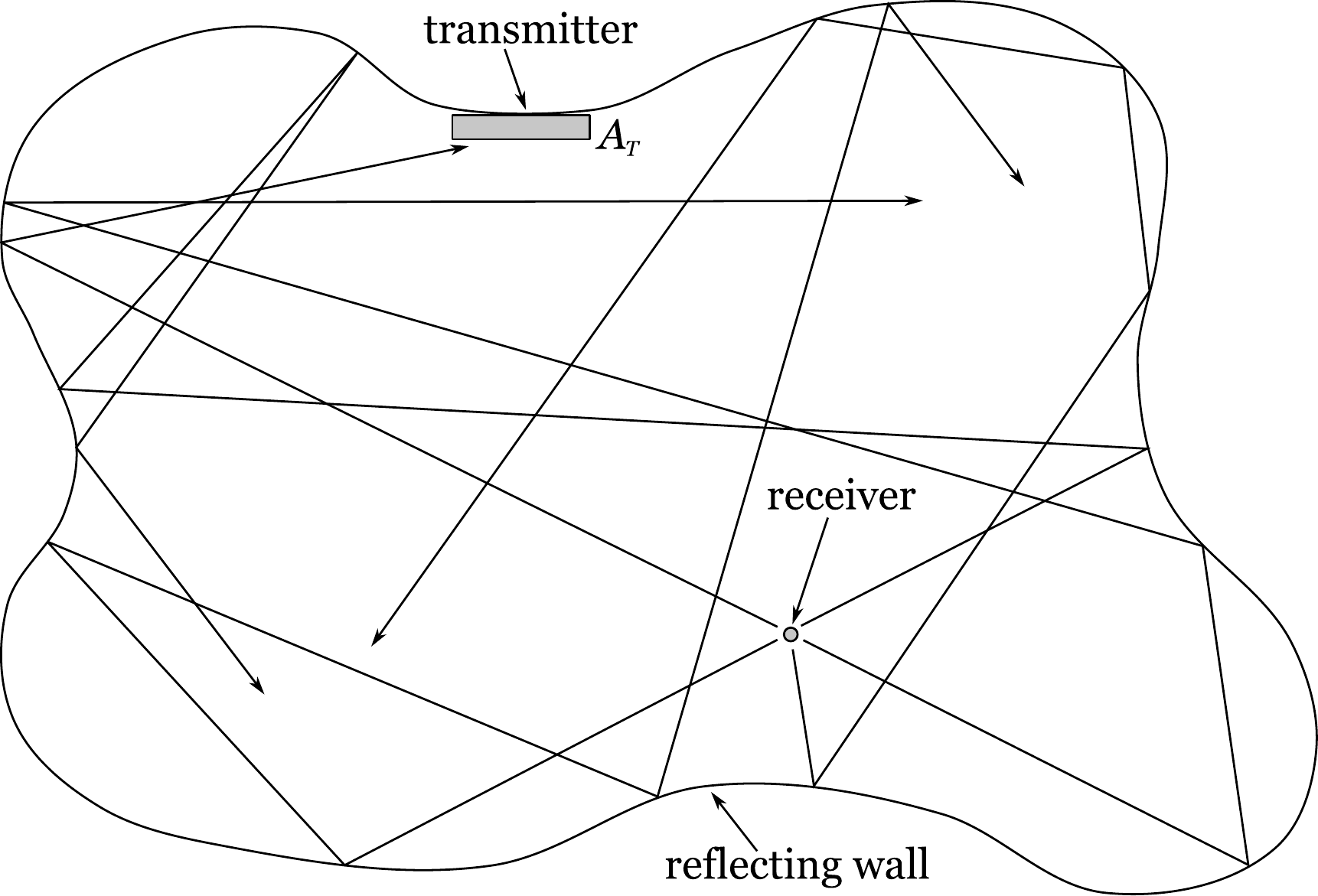}
\caption{A transmitter and a receiver inside a room of arbitrary shape. The arrows represent radiation spreading out from the receiver and scattering from the walls.}
\label{fig:arbitrary_room}
\end{center}
\end{figure}

The form of equation \ref{eqn:box_reflection_efficiency} suggests an alternate method for calculating the efficiency of power transfer in a room with reflecting walls which is applicable to arbitrary arrangements of a transmitter and a receiver in a room of arbitrary shape. Consider the arrangement shown in Figure \ref{fig:arbitrary_room}. As before, we may calculate the efficiency using reciprocity, in which the receiver is treated as the source of power. Let $\rho$ represent the power reflection coefficient, and $\alpha=1-\rho$ represent the power absorption coefficient of the walls. Both the absorption and reflection coefficients may be functions of position. The power incident on the walls may be reflected diffusely, specularly, or in some combination thereof. 

Assume that the surface area of the transmitter, $\mathcal{A}_T$, and the absorption coefficient, $\alpha$, are sufficiently small that the radiation is capable of undergoing many reflections before being absorbed. If this is the case, then the system will be ergodic, and the room will become filled with a homogeneous and locally isotropic bath of radiation.

Because the same power density, $S$, is incident on all surfaces in the room, the power which is absorbed by the transmitter, $P_T$, is:
\begin{equation}
P_T = S \mathcal{A}_T
\end{equation}
The power absorbed by the walls, $P_W$, is given by the integral:
\begin{equation}
P_W = \int d\mathcal{A}_W \alpha S
\end{equation}
where the integral is taken over the two-dimensional surface of the wall. We may define an effective absorbing area, $\mathcal{A}_A$, to be:
\begin{equation}
\mathcal{A}_A \equiv \int d\mathcal{A}_W \alpha
\end{equation}
The efficiency of power transfer due to the reflections from the walls, $\eta_{\textrm{refl.}}$, is therefore:
\begin{equation}
\eta_{\textrm{refl.}} = \frac{P_T}{P_W + P_T} = \frac{\mathcal{A}_T}{\mathcal{A}_A + \mathcal{A}_T}
\label{eqn:reflection_efficiency_general_case}
\end{equation}
Here we find that the formula for the efficiency, $\eta_{\textrm{refl.}}$, is the same as that derived in equation \ref{eqn:box_reflection_efficiency} for a cubical room, except that equation \ref{eqn:reflection_efficiency_general_case} is applicable to the general case of a room of arbitrary shape.
If the efficiency of power transfer by the direct path is $\eta_{\textrm{direct}}$, then the overall system efficiency, $\eta_{\textrm{room}}$, is:
\begin{equation}
\eta_{\textrm{room}} = \eta_{\textrm{direct}} + \eta_{\textrm{refl.}} -  \eta_{\textrm{direct}} \eta_{\textrm{refl.}}
\end{equation}
As before, we see that the efficiency approaches 100\% as $\alpha$ approaches 0.

\section{The Amplitude and Phase of the Transmitter Array}
\label{sec:amplitude_and_phase}

It may seem, at first glance, that the assumption of ergodicity made in the previous section should be true in reverse, i.e.\ that the room is filled with a homogeneous and locally isotropic bath of radiation when power is sent from the transmitter to the receiver. A simple calculation, however, proves that this cannot be the case.

For the sake of argument, let us assume that the transmitter emits radiation with an arbitrary gain and phase pattern, such that the radiation scatters randomly from the walls and fills the room with a homogeneous and isotropic bath of radiation. Let $S$ denote the power density incident on all of the surfaces in the room. The power re-absorbed by the transmitter, $P_T$, is:
\begin{equation}
P_T = S \mathcal{A}_T
\end{equation}
the power absorbed by the walls, $P_W$, is:
\begin{equation}
P_W = S \mathcal{A}_A
\end{equation}
and the power absorbed by the receiver, $P_R$, is:
\begin{equation}
P_R = S \overline{\mathcal{A}}_R
\end{equation}
where $\overline{\mathcal{A}}_R = \lambda^2 / (2 \pi)$ is the effective absorbing area of the receiver averaged over all directions.\footnote{See Appendix \ref{sec:antenna_average_effective_area}.} The net power emitted by the transmitter is equal to the total power absorbed by all surfaces minus the power re-absorbed by the transmitter:
\begin{equation}
P_{\textrm{net}} = (P_T + P_W + P_R) - P_T = P_W + P_R
\end{equation}
The efficiency is:
\begin{equation}
\eta = \frac{P_R}{P_{\textrm{net}}} = \frac{P_R}{P_W + P_R} = \frac{\overline{\mathcal{A}}_R}{\mathcal{A}_A + \overline{\mathcal{A}}_R}
\label{eqn:ergodic_transmitter}
\end{equation}
This result is similar to equation \ref{eqn:reflection_efficiency_general_case}, except that the surface area of the transmitter, $\mathcal{A}_T$, has been replaced by the effective area of the receiver, $\overline{\mathcal{A}}_R$. If the surface area of the transmitter is much larger than the effective area of the receiver, the efficiency predicted by equation \ref{eqn:ergodic_transmitter} is much less than that predicted by equation \ref{eqn:reflection_efficiency_general_case}, even though these two efficiencies are supposed to be equal according to the Reciprocity Theorem.

In order for the reciprocity theorem to hold, the receiver must be surrounded by a power density, $S_R$, which is enhanced relative to $S$ by the factor:
\begin{equation}
\frac{S_R}{S} = \frac{\mathcal{A}_T}{\overline{\mathcal{A}}_R}
\end{equation}
We therefore see that the power density within the room cannot be homogeneous and locally isotropic when power is sent from the transmitter to the receiver.

A resolution to this paradox is suggested by the analysis from section \ref{sec:reflections}, in which we saw that the presence of the reflecting wall enhanced the efficiency of the power transfer only when the amplitudes and phases of the outgoing beams from the transmitter were chosen so as to produce constructive interference in the vicinity of the receiver. The radiation pattern from the transmitter antenna must therefore be carefully chosen so as to produce constructive interference at the location of the receiver.

\begin{figure}[htb]
\begin{center}
\includegraphics[width=0.5\textwidth]{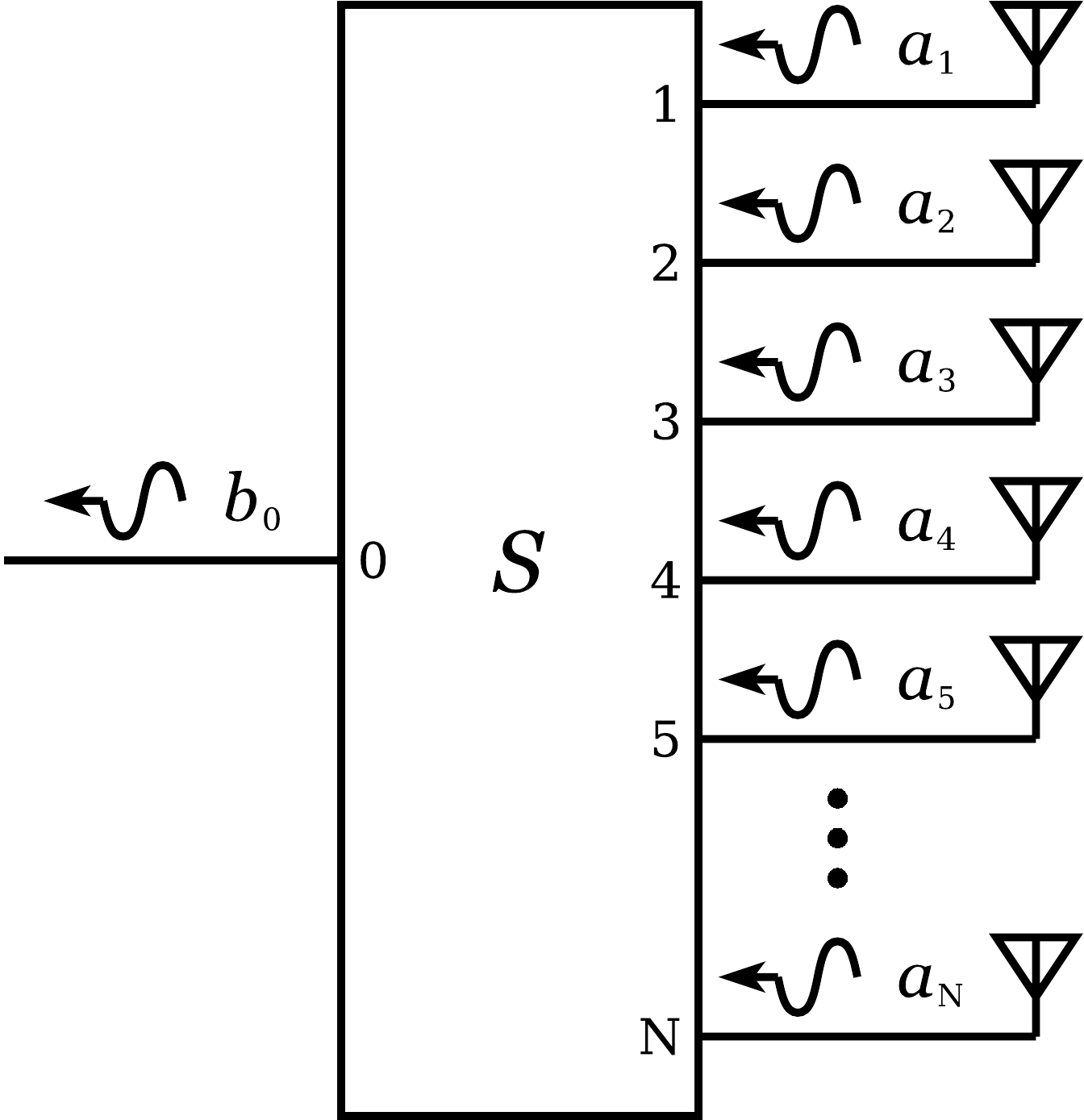}
\caption{A transmitter consisting of an array of small antennas connected to a multi-port splitter with S-matrix $S_{mn}$. The network combines the power from the array of $N$ antennas, and sends the power out through port 0.}
\label{fig:antenna_array}
\end{center}
\end{figure}

Consider the schematic for a transmitter antenna shown in Figure \ref{fig:antenna_array}. The transmitter consists of an array of $N$ small antennas connected to a multi-port splitter with S-matrix $S_{mn}$. Let the receiver be treated as the source of the power. Assume that each of the antennas in the transmitter array is matched to its characteristic impedance. Let $a_n$ denote the complex amplitude of the wave received by the $n$th antenna.

For a given set of received amplitudes, $a_n$, the S-parameters may always be chosen such that the reflection coefficient at each port, $S_{nn}$, is zero, and all of the power from the antennas is combined and sent out through port 0, which is indicated by the outgoing wave of amplitude $b_0$ in Figure \ref{fig:antenna_array}.

Because the system is linear, we may re-scale the complex amplitudes, $a_n$, such that:
\begin{equation}
\sum_{n=1}^{N} |a_n|^2 = 1 \ \ \ \ \textrm{and} \ \ \ \ b_0 = 1
\end{equation}
All of the above conditions are met by the following choice of S-parameters:
\begin{eqnarray}
S_{00} &=& 0\nonumber\\
S_{0n} &=& a_n^\ast \ \ , \ \ 1 \leq n \leq N\nonumber\\
S_{mn} &=& 0\ \ , \ \ 1 \leq n \leq N \ \ , \ \ 1 \leq m \leq N
\end{eqnarray}
By reciprocity, the following must also be true:
\begin{equation}
S_{n0} = a_n^\ast  \ \ , \ \ 1 \leq n \leq N
\end{equation}
Therefore, when a wave of unit amplitude is sent into port 0, the complex amplitude, $b_n$, of the $n$th antenna in the array is:
\begin{equation}
b_n = S_{n0} = a_n^\ast
\end{equation}
which is the conjugate of the re-scaled complex amplitude of the signal received by the $n$th antenna when the receiver was treated as the source of power.

This immediately suggests a method for determining the correct choice of complex amplitudes, $b_n$, so as to maximize the power absorbed by the receiver: If the receiver first sends out a signal, the amplitude and phase of each antenna in the transmitter array may be measured \cite{Leabman2014}. In order to send power back to the receiver with the same efficiency, the transmitter need only drive each antenna in the array with a complex amplitude proportional to the conjugate of the amplitude which that antenna received.

\section{Multiple Receivers}

Wireless power in a modern domestic setting finds its greatest use when it is capable of simultaneously providing power to multiple devices. For this reason, it is of interest to investigate how the method for wireless power transfer described in the previous two sections may be applied to the case in which a multitude of receivers are to be powered simultaneously by a single transmitter within an enclosed space.

\begin{figure}[htb]
\begin{center}
\includegraphics[width=0.8\textwidth]{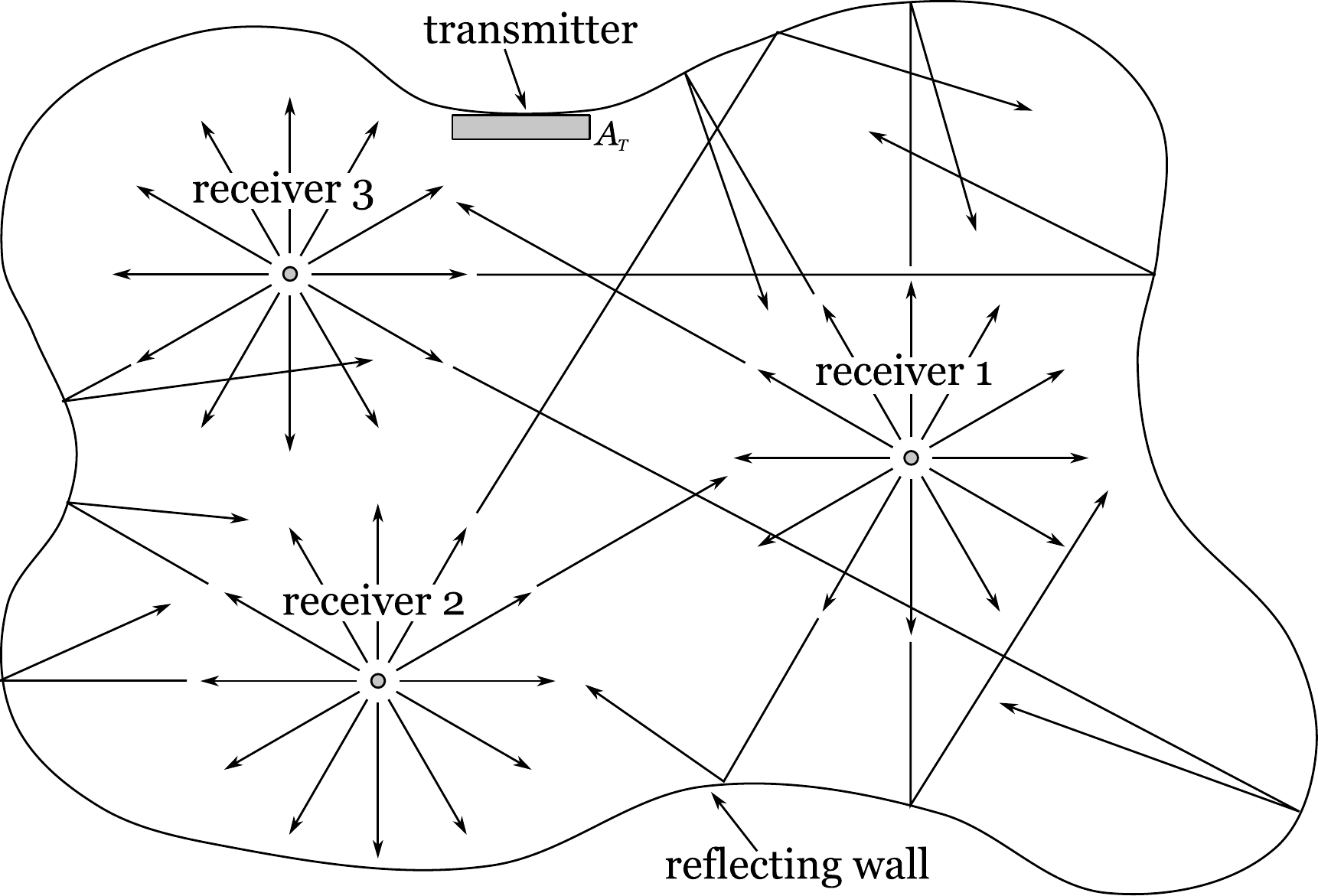}
\caption{A transmitter and multiple receivers inside a room of arbitrary shape. The arrows represent radiation spreading out from the receivers and scattering from the walls.}
\label{fig:arbitrary_room_multiple_receivers}
\end{center}
\end{figure}

Such a configuration is depicted in Figure \ref{fig:arbitrary_room_multiple_receivers}. As before, the efficiency of power transfer may be calculated using reciprocity. Assume that a total power, $P_R$, is split among the multiple receivers and radiated into the room. As before, the overall efficiency will be equal to the fraction of this power absorbed by the transmitter.

The signal received by the transmitter will be a linear combination of the signals generated by each of the receivers. The total power absorbed by the transmitter is the integral of the squared amplitude of this sum over the surface area of the transmitter. At some locations on the surface of the transmitter the signals interfere constructively, while at other locations they interfere destructively. On average, unless the receivers are separated by a distance less than a wavelength, the received power is the incoherent sum of the power from each of the signals.

Therefore, assuming, as before, that the system is ergodic, the fraction of power absorbed by the transmitter due to reflections from the walls is the given by equation \ref{eqn:reflection_efficiency_general_case}. The overall system efficiency is therefore independent of the number of receivers. If each receiver needs a fixed amount of power, then the total power emitted by the transmitter must be proportional to the number of receivers.

In order for the transmitter to transfer power to each of the receivers, the amplitude and phase of each of the antennas in the transmitter array must be chosen so as to produce constructive interference in the vicinity of each of the receivers. Because the the efficiency is independent of the phase of each receiver, the outgoing radiation from the transmitter can be composed of a linear combination of signals designed to create constructive interference in the vicinity of each receiver individually, where the overall phase of each component of the linear combination can be chosen arbitrarily.

The complex amplitude of each element of the transmitter array required to produce a signal which concentrates power in the vicinity of a receiver can be determined, as before, by finding the conjugate of the the amplitude and phase received by each antenna element when that receiver is the source of power.

\section{An Upper Limit on Undesired Absorption}

The previous sections provide formulas for the overall efficiency, $\eta$, of power transfer from a transmitter to one or more wireless receivers within an enclosed space. Given a total desired power, $P_R$, to be received by all of the receivers, the power, $P_T$, which the transmitter must emit is given by:
\begin{equation}
P_T = \frac{P_R}{\eta}
\end{equation}
We see that regardless of the value of the efficiency, $\eta$, it is always possible to satisfy the power requirements of the receivers by choosing a transmitter with a sufficiently high power output.

However, in a domestic setting, the transmitted power, $P_T$, cannot be increased without bound. Human beings will necessarily be present within the enclosed area. At some point, the maximum permissible radiation exposure limit for these human beings will be exceeded.

For this reason, it is of interest to calculate the maximum power which can be safely delivered to the receivers without exceeding an upper limit on the undesired absorption of power by certain bodies within the room. Consider the case in which the room contains an absorbing body with absorbing area, $\mathcal{A}_B$, whose power absorption must be limited to be no more than a power, $P_{B\textrm{max}}$. Neglecting the direct path, the efficiency, $\eta$, of power transfer from the transmitter to one or more receivers is:
\begin{figure}[t]
\begin{center}
\includegraphics[width=0.8\textwidth]{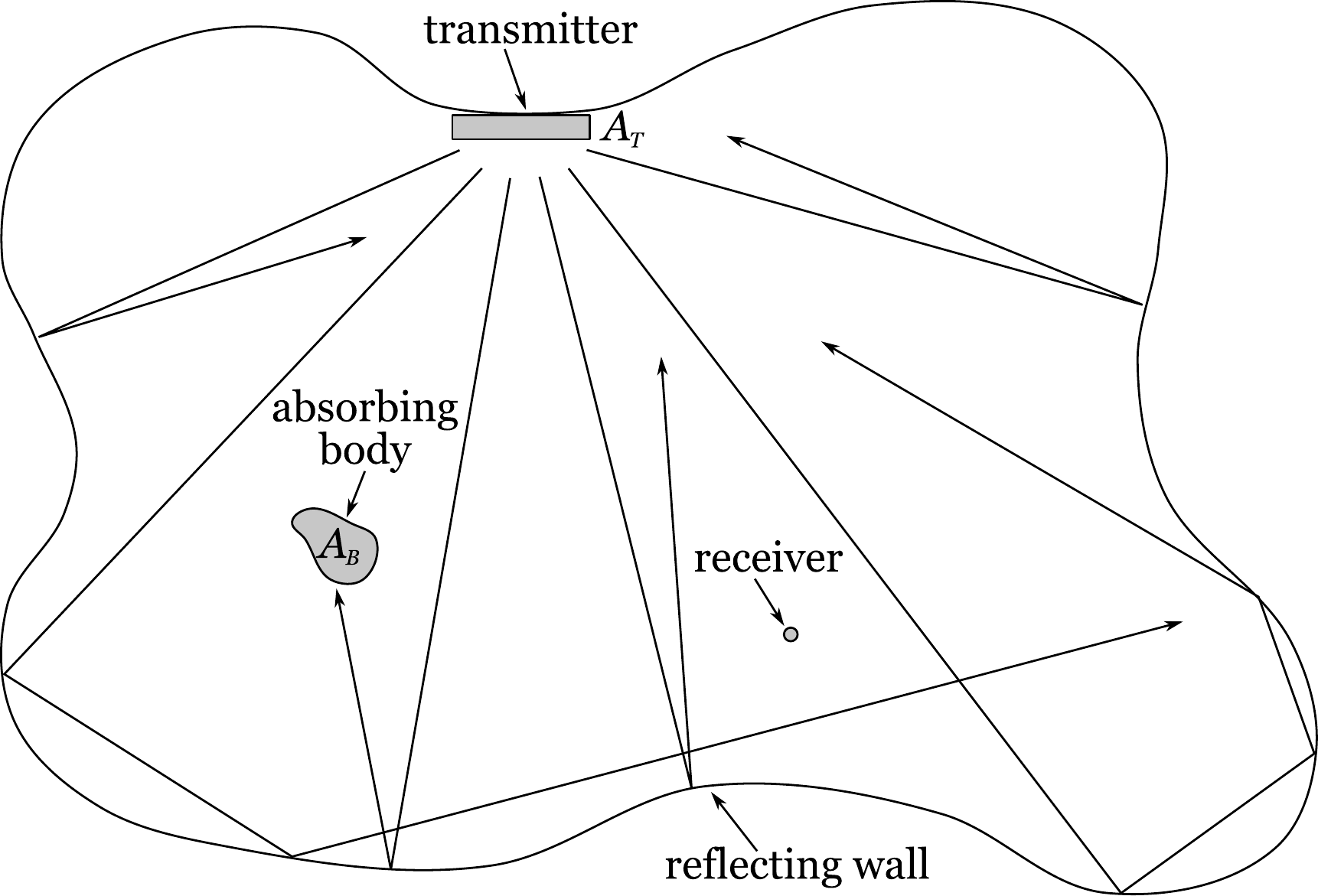}
\caption{A room containing a transmitter, receiver, and an absorbing body with effective absorbing area $\mathcal{A}_B$.}
\label{fig:arbitrary_room_absorbing_body}
\end{center}
\end{figure}
\begin{equation}
\eta = \frac{\mathcal{A}_T}{\mathcal{A}_A + \mathcal{A}_B + \mathcal{A}_T}
\end{equation}
where $\mathcal{A}_A$ is the effective absorbing area of all other surfaces in the room. Assume that the transmitter emits a total power $P_T$. The amount of power absorbed by all surfaces other than the receivers is:
\begin{equation}
(1- \eta) P_T = \frac{\mathcal{A}_A + \mathcal{A}_B}{\mathcal{A}_A + \mathcal{A}_B + \mathcal{A}_T} P_T
\end{equation}
We wish to determine the fraction of this power absorbed by the surface of the absorbing body, $\mathcal{A}_B$. Because the radiation in the room is not homogeneous when the power is sent from the transmitter to the receivers, we cannot necessarily assume that the surface of the absorbing body, $\mathcal{A}_B$, is exposed to the same power density as the rest of the absorbing area, $\mathcal{A}_A$. However, if the absorbing body is free to move about the room, we may assume that, averaged over all configurations, the radiation is not preferentially absorbed by any particular absorbing surface. Therefore, we may estimate the power absorbed by the surface of the absorbing body averaged over all configurations, $\overline{P}_B$, to be:
\begin{equation}
\overline{P}_B = \frac{\mathcal{A}_B}{\mathcal{A}_A + \mathcal{A}_B} (1-\eta) P_T = \frac{\mathcal{A}_B}{\mathcal{A}_A + \mathcal{A}_B + \mathcal{A}_T} P_T
\label{eqn:absorbing_body_power}
\end{equation}
Using equation \ref{eqn:absorbing_body_power}, we can find a relation between the power transferred to the receivers, $P_R$, and the average power absorbed by the absorbing body, $\overline{P}_B$:
\begin{equation}
P_R = \eta P_T = \frac{\mathcal{A}_T}{\mathcal{A}_A + \mathcal{A}_B + \mathcal{A}_T} \frac{\mathcal{A}_A + \mathcal{A}_B + \mathcal{A}_T}{\mathcal{A}_B} \overline{P}_B = \frac{\mathcal{A}_T}{\mathcal{A}_B} \overline{P}_B
\end{equation}
Because the power absorbed by the absorbing body is limited to be no greater than $P_{B\textrm{max}}$, we may place the following upper limit on the power transferred to the receivers:
\begin{equation}
P_R \leq \frac{\mathcal{A}_T}{\mathcal{A}_B} P_{B\textrm{max}}
\label{eqn:max_received_power}
\end{equation}

For the case where the absorbing body is a human being, $\mathcal{A}_B$ is equal to the surface area of the human body weighted by its power absorption coefficient at the frequency of interest. The maximum absorbed power, $P_{B\textrm{max}}$, is set by various regulatory agencies. Note that both $\mathcal{A}_B$ and $P_{B\textrm{max}}$ are directly proportional to the number of humans in the room, so the upper limit on $P_R$ is independent of this number.

Note that equation \ref{eqn:max_received_power} may also be rearranged in the following way:
\begin{equation}
\frac{P_R}{\mathcal{A}_T} \leq \frac{P_{B\textrm{max}}}{\mathcal{A}_B}
\end{equation}
in which it becomes a comparison of two power densities. Here we see that the ratio of the total received power, $P_R$, to the surface area of the transmitter, $\mathcal{A}_T$, must be less than or equal to the maximum permissible average power density on the surface of the absorbing body.

\section{Controlled Destructive Interference}

Given the fact that an antenna array can produce regions of constructive interference within an enclosed space, it is natural to ask whether a similar technique to that described in section \ref{sec:amplitude_and_phase} might allow for the creation of regions of destructive interference \cite{Leabman2014}. The creation of such regions would be useful, for example, in minimizing the radiation exposure of human beings present within the enclosed space.

Consider the situation depicted in Figure \ref{fig:arbitrary_room_absorbing_body} in which an absorbing body is placed in a room containing a transmitter and one or more receivers. Suppose we wish to emit radiation from the transmitter in such a way as to exactly cancel the fields present on the surface of the absorbing body, thereby enveloping it in a region of destructive interference, while at the same time transferring a non-zero power to the receivers.

In order to compute the necessary field pattern, let us first imagine covering the surface of the absorbing body with an array of $M$ patch antennas. Let us assume that the transmitter consists of an array of $N$ patch antennas. The radiation received by the absorbing body is given by the linear relation:
\begin{equation}
\texttt{b} = \texttt{S} \, \texttt{a}
\end{equation}
where $\texttt{b}$ is an $M$-row column vector containing the complex amplitudes of the antenna elements in the array on the absorbing body, $\texttt{a}$ is an $N$-row column vector containing the complex driving amplitudes of the antenna elements in the transmitter array, and $\texttt{S}$ is an $M \times N$ matrix.

Let $\texttt{a}_1$ be a signal which creates a constructive interference pattern in the vicinity of the receiver. The resulting fields, $\texttt{b}_1$, received by the antennas on the absorbing body are:
\begin{equation}
\texttt{b}_1 = \texttt{S} \, \texttt{a}_1
\end{equation}
We wish to find a signal, $\texttt{a}_2$, which may be added to the signal, $\text{a}_1$, such that:
\begin{eqnarray}
\texttt{b}_2 &=& \texttt{S} \, \texttt{a}_2\nonumber\\
\texttt{S} \, (\texttt{a}_1 + \texttt{a}_2 ) &=& \texttt{b}_1 + \texttt{b}_2 = 0
\end{eqnarray}
A potential solution is:
\begin{equation}
\texttt{a}_2 = -\texttt{S}^\dagger \left( \texttt{S} \texttt{S}^\dagger \right)^{-1} \texttt{S} \, \texttt{a}_1
\label{eqn:destructive_interference_solution}
\end{equation}
where $\texttt{S}^\dagger$ is the Hermetian conjugate of $\texttt{S}$. In order for such a solution to exist, it is necessary that the $M \times M$ matrix, $\texttt{S} \texttt{S}^\dagger$, be invertible. Note that the rank of the matrix, $\texttt{S} \texttt{S}^\dagger$, can be no greater than $N$. Therefore, the solution given in equation \ref{eqn:destructive_interference_solution} cannot exist unless $N\geq M$.

Assuming that the antennas in the transmitter array have the same spacing as those on the surface of the absorbing body, the results above imply the following necessary constraint on the surface area of the transmitter, $\mathcal{A}_T$, relative to the total surface area of the absorbing body, $\mathcal{A}_{B\textrm{total}}$, in order for it to be possible to create the region of destructive interference:
\begin{equation}
\mathcal{A}_T \geq \mathcal{A}_{B\textrm{total}}
\end{equation}

\section{Conclusions}

The results above show that reflections from the walls of an enclosed space have the capacity to greatly enhance the efficiency of radiative wireless power transfer in the limit where the walls are highly reflective. In this limit, as the distance between the transmitter and receiver is increased, the efficiency becomes independent of the position of the receiver and instead depends only on the ratio of the surface area of the transmitter to the effective absorbing area of the room.

In order for the efficiency to be optimized, the transmitter must create a region of constructive interference in the vicinity of the receiver. A simple method was described for determining the field pattern necessary for achieving this effect using a signal sent in reverse from the receiver to the transmitter.

If the enclosed space contains an absorbing body with an upper limit on its allowed power absorption, such as a human being, there exists an upper limit on the power which may be transferred to the receivers. This upper limit depends only on the surface area of the transmitter and the maximum permissible average power density on the surface of the absorbing body.

This upper limit on the received power may be circumvented if the radiation from the transmitter is precisely tuned to produce destructive interference on the entire surface of the absorbing body. However, in order to produce the region of destructive interference, the surface area of the transmitter must be greater than or equal to the surface area of the absorbing body.

\newpage

\appendix

\section{The Average Effective Area of an Antenna}
\label{sec:antenna_average_effective_area}

Consider an ideal single-port antenna with gain pattern, $G$. Assume that the antenna is surrounded by a locally isotropic bath of radiation with radiance, $\mathcal{R}$. The total power absorbed by the antenna, $P_{\textrm{ant.}}$, is:
\begin{equation}
P_{\textrm{ant.}} = \frac{1}{2} \int d\Omega \, \mathcal{R} \mathcal{A}_{\textrm{ant.}}
\label{eqn:antenna_power}
\end{equation}
where $\mathcal{A}_{\textrm{ant.}} = G \lambda^2/(4\pi)$ is the effective area of the antenna, and the factor of $\nicefrac{1}{2}$ accounts for the two polarization directions. Substituting this into equation \ref{eqn:antenna_power}, we get:
\begin{equation}
P_{\textrm{ant.}} = \frac{1}{2} \int d\Omega \, \mathcal{R} \, G \frac{\lambda^2}{4 \pi} = \frac{1}{2} \mathcal{R} \lambda^2
\end{equation}
We may compare this to the power, $P_{\textrm{surf.}}$, absorbed by a perfectly-absorbing surface of area $\mathcal{A}$ exposed on one side to the same isotropic bath of radiation:
\begin{equation}
P_{\textrm{surf.}} = \int d\Omega \, \mathcal{R} \mathcal{A} \cos{\theta}
\end{equation}
where $\theta$ is the angle of incidence of the incoming radiation, and where the angular integral, $\int d\Omega$, is taken only over one hemisphere. The result is:
\begin{equation}
P_{\textrm{surf.}} = \pi \mathcal{R} \mathcal{A}
\end{equation}
If we set $P_{\textrm{surf.}} = P_{\textrm{ant.}}$, we find that the power absorbed by the antenna is equivalent to that absorbed by a surface of area, $\overline{\mathcal{A}}$, given by:
\begin{equation}
\overline{\mathcal{A}} = \frac{\lambda^2}{2 \pi}
\end{equation}
which is true regardless of the gain pattern of the antenna.

\end{document}